\documentclass[showpacs,amssymb,aps]{revtex4}

\usepackage{graphicx}
\begin{document}

\title{Kontsevich product and gauge invariance}  

\author{Ashok Das$^{a}$ and Josif Frenkel$^{b}$}
\affiliation{$^{a}$ Department of Physics and Astronomy,
University of Rochester,
Rochester, NY 14627-0171, USA}
\affiliation{$^{b}$ Instituto de F\'{\i}sica, Universidade de S\~ao
Paulo, S\~ao Paulo, SP 05315-970, BRAZIL}

\bigskip
%\date{}

\begin{abstract}

We analyze the question of $U_{\star} (1)$ gauge invariance in a flat 
non-commutative space where the parameter of non-commutativity,
$\theta^{\mu\nu} (x)$, is a
local function satisfying Jacobi identity (and thereby leading to an
associative Kontsevich product). We show
that in this case, both gauge transformations as well as the
definitions of covariant derivatives have to modify so as to have a
gauge invariant action. We work out the gauge invariant actions for
the matter fields in the fundamental and the adjoint representations up to
order $\theta^{2}$ while we discuss the gauge invariant Maxwell theory
up to order $\theta$. We show that despite the modifications in the
gauge transformations, the covariant derivative and the field
strength, Seiberg-Witten map continues to hold for this theory. In
this theory, translations do not form a subgroup of the gauge
transformations (unlike in the case when $\theta^{\mu\nu}$ is a
constant) which is reflected in the stress tensor not being conserved. 

\end{abstract}


\maketitle

\section{Introduction}

Non-commutativity of space-time coordinates have been by now studied
exhaustively from various points of view
\cite{snyder,connes,douglas,douglas1,schomerus,witten}. It
arises naturally  in the
quantization of open strings and membranes attached to $D$-branes in
the presence of background fields \cite{sj,chu,das}. The
non-commutativity is related to the
background fields so that if the background fields are constant, one
obtains the more familiar case where the parameter of
non-commutativity $\theta^{\mu\nu}$ is a constant. In such a case, the
standard multiplication of functions is replaced by the
Gr\"{o}newold-Moyal product \cite{moyal}. On the other hand, if the background
fields depend on space-time coordinates, then, one expects the
parameter of non-commutativity to be a local function. Theories with
a local parameter of non-commutativity have not been studied as
vigorously. In this case, the multiplication of functions is replaced
by the Kontsevich product \cite{kontsevich} and there are two natural
cases  that can 
arise. Namely, the parameter of non-commutativity $\theta^{\mu\nu}
(x)$ satisfies the Jacobi identity in which case the Kontsevich
product is associative. The second possibility is that
$\theta^{\mu\nu} (x)$ does not satisfy the Jacobi identity leading to
a non-associative Kontsevich product. The first case corresponds
to the embedding of a curved $D$-brane in a flat background while the second
arises for a curved brane embedded in a curved background (with the
non-associativity related to the curvature of the background)
\cite{cornalba}. In the case where
$\theta^{\mu\nu}$ is local and satisfies Jacobi identity, there exists
so far only a single analysis of a model, namely, the
Cattaneo-Felder model which involves the study of a boundary conformal
field theory \cite{cattaneo}. In this
paper, we would like to extend such a possibility to the case of gauge
theories. 

This is also important from the point of view of studying the
properties of non-commutative field theories (independent of their
origin). By now, non-commutative field theories in flat
space-time with $\theta^{\mu\nu}$ constant have been studied
extensively and various interesting properties have been
noted \cite{douglas1}. However, eventually one would like to study the
properties of
such theories in a curved background (possibly including
non-commutative gravity \cite{chamseddine}). In such a case, it is clear that
$\theta^{\mu\nu}$ can no longer be considered a constant. As a first
attempt at studying such theories, it would be interesting to study
the behavior of a field theory in a flat non-commutative space-time
where $\theta^{\mu\nu}$ is a local function. In fact, it is even more
interesting to study the question of gauge invariance in such a
case. With that in mind, we have chosen to study a $U_{\star} (1)$
gauge theory with matter in the fundamental as well as in the adjoint
representations \cite{hayakawa} and we find many interesting features from our
analysis of such theories.

Let us recall that when $\theta^{\mu\nu}$ is local, the star product
is given by the Kontsevich product \cite{kontsevich,cornalba,cattaneo}
\begin{equation}
f\star g = fg + \frac{i}{2} \theta^{\mu\nu} \partial_{\mu}f
\partial_{\nu}g - \frac{1}{8} \theta^{\mu\nu} \theta^{\lambda\rho}
\partial_{\mu}\partial_{\lambda} f \partial_{\nu}\partial_{\rho} g -
\frac{1}{12} \theta^{\lambda\rho}\partial_{\rho}\theta^{\mu\nu}
\left(\partial_{\lambda}\partial_{\mu} f \partial_{\nu} g -
\partial_{\mu} f \partial_{\lambda}\partial_{\nu} g\right) + \cdots\,
,\label{kproduct}
\end{equation}
where $\cdots$ represent higher order terms in $\theta$. It is clear
from (\ref{kproduct}) that we can identify 
\begin{equation}
\left[x^{\mu} , x^{\nu}\right] = x^{\mu}\star x^{\nu} - x^{\nu}\star
x^{\mu} = i \theta^{\mu\nu} (x)\, ,
\end{equation}
and if $\theta^{\mu\nu} (x)$ satisfies the Jacobi identity,
\begin{equation}
\theta^{\mu\alpha} \partial_{\alpha} \theta^{\nu\lambda} +
\theta^{\nu\alpha}\partial_{\alpha} \theta^{\lambda\mu} +
\theta^{\lambda\alpha} \partial_{\alpha} \theta^{\mu\nu} = 0\,
,\label{jacobi} 
\end{equation}
then, the product (\ref{kproduct}) is associative. Throughout this
paper, we will restrict ourselves to such a case. It is also clear
from (\ref{kproduct}) that we can represent
\begin{equation}
i \theta^{\mu\nu} (x) \partial_{\nu} f (x) = \left[x^{\mu}, f\right]\,
.
\end{equation}
If $\theta^{\mu\nu} (x)$ has an inverse, this can even be inverted to
write
\begin{equation}
\partial_{\mu} f (x) = - i (\theta^{-1})_{\mu\nu} (x) \left[x^{\nu},
  f\right] \neq \left[-i (\theta^{-1})_{\mu\nu} x^{\nu},f\right]\,
  ,\label{derivative}
\end{equation}
where the last relation holds only for the case when $\theta^{\mu\nu}$ is a
constant.

Unlike the more
studied case of constant $\theta^{\mu\nu}$, the products inside an
integral no longer satisfy cyclicity when the parameter of
non-commutativity is a local function and as a result, the analysis of
such theories is a bit more involved. We note here that when
$\theta^{\mu\nu}$ is a local function, it can be thought of as a
genuine Lorentz tensor. On the other hand, we do know that unitarity
in a non-commutative theory is violated unless $\theta^{0i} =
0$ \cite{gomis}. Presumably, one can make such a choice by going to a
particular reference 
frame. However, for purposes of studying gauge invariance properties
of the action at the tree level, we do not concern ourselves with
this question in this paper. This certainly is an important question
which deserves further study and
we will report on this in the future. In this paper, we assume that
$\theta^{\mu\nu} (x)$ is a genuine tensor (although that really does
not enter into our analysis at all).

The organization of our paper is as follows. In section {\bf II}, we
discuss  how matter (scalar fields) in the fundamental as well as in
the adjoint representations can be coupled to the photon in a
$U_{\star} (1)$ invariant manner. This necessitates a modification of
the covariant derivative as well as the gauge transformation for the
gauge field. We demonstrate how this can be achieved systematically up
to order $\theta^{2}$ in the matter sector. The definition of the
field strength also changes as a consequence and we determine the
gauge invariant action for the Maxwell theory up to order $\theta$
(since it is much more involved than the matter sector because of the
Lorentz indices, but the
procedure is clear). One of the surprising outcomes of this analysis
is that $\theta^{\mu\nu} (x)$ does not transform under the gauge
transformation (although, naively, one would have expected it to
transform in the adjoint representation). We note that all the
modifications that we find vanish when $\theta^{\mu\nu}$ is a constant
reducing to the conventional $U_{\star} (1)$ gauge invariance studied
in the literature \cite{hayakawa}. In section
{\bf III}, we
show that in spite of these modifications, the Seiberg-Witten map
\cite{witten} 
between the non-commutative and the commutative theories continues to
hold. This is surprising and suggests some deeper meaning of the map
that we have not studied further. We show the equivalence of the
equations of motion (non-commutative and commutative) as well as the stress
tensors of the Maxwell theory under the map. Furthermore, we show that
in the present case (unlike in the case of a constant
$\theta^{\mu\nu}$ \cite{frenkel}), the stress tensor is not
conserved. This is traced
to the fact that when $\theta^{\mu\nu} (x)$ is a local function, it
can be thought of as an external field which violates translation
invariance. In fact, since $\theta^{\mu\nu} (x)$ is inert under a
gauge transformation, while
translation invariance requires it to transform, it follows that
in this case, translations do not form a subgroup of the $U_{\star}
(1)$ gauge transformations (as is the case for constant
$\theta^{\mu\nu}$ \cite{douglas1}). We close with a brief summary in
section {\bf IV}.

\section{Gauge invariant actions}

In this section, we will construct actions invariant under $U_{\star}
(1)$ gauge transformations. Let us start with the action for a complex
scalar field (which would represent matter in the fundamental
representation), which conventionally has the form
\begin{equation}
S_{\rm fund} = \int \mathrm{d}x\, \left((D_{\mu}\phi)^{*}\star (D^{\mu}\phi) -
m^{2} \phi^{*}\star \phi\right)\, .\label{action1}
\end{equation}
We have left the dimensionality of space-time arbitrary since that
does not enter into our analysis. In the more familiar case of a
constant $\theta^{\mu\nu}$, the covariant derivative has the form
\cite{hayakawa}
\begin{equation}
D_{\mu}\phi = \partial_{\mu}\phi - i A_{\mu}\star\phi\,
,\label{covariant0} 
\end{equation}
and the action is invariant under the infinitesimal gauge transformations
\begin{equation}
\delta \phi = i \epsilon\star \phi,\qquad \delta A_{\mu} =
\partial_{\mu}\epsilon - i \left[A_{\mu},\epsilon\right]\,
,\label{tfn0} 
\end{equation}
where $\epsilon (x)$ represents the infinitesimal parameter of gauge
transformations. When $\theta^{\mu\nu}$ becomes a local function, the
action in (\ref{action1}) with (\ref{covariant0}) is no longer
invariant under the gauge transformations (\ref{tfn0}). In this case,
we have to systematically determine the modifications necessary in the
definitions of the covariant derivative and the gauge transformations
under which the action (\ref{action1}) will be invariant. We will
demonstrate how this can be done up to order $\theta^{2}$ in this
theory. 

To begin with, let us note that when $\theta^{\mu\nu}$ is local,
\begin{equation}
\int \mathrm{d}x\, A (x)\star B (x) \neq \int \mathrm{d}x\, B (x)
\star A (x)\, .
\end{equation}
However, form the definition of the product in (\ref{kproduct}), we
note that
\begin{equation}
\left(A(x)\star B (x)\right)^{*} = B^{*} (x)\star A^{*} (x)\,
,\label{identity1} 
\end{equation}
so that if we can find a definition of the covariant derivative as
well as gauge transformations such that
\begin{equation}
\delta \phi = i \alpha (x)\star \phi,\qquad \delta (D_{\mu}\phi) = i
\beta (x)\star (D_{\mu}\phi)\, ,\label{req1}
\end{equation}
for real functions $\alpha (x), \beta (x)$, then, action
(\ref{action1}) will be gauge invariant.

To systematically determine these modifications, let us represent
\begin{eqnarray}
\delta \phi & = & i \epsilon (x)\star \phi (x) + P (x)\, ,\nonumber\\
\noalign{\vskip 4pt}%
\delta A_{\mu} & = & \partial_{\mu}\epsilon (x) - i
\left[A_{\mu},\epsilon\right] + Y_{\mu} (x)\, ,\nonumber\\
\noalign{\vskip 4pt}%
D_{\mu}\phi (x) & = & \partial_{\mu}\phi (x) - i A_{\mu}\star \phi +
Z_{\mu} (x)\, ,\label{modif1}
\end{eqnarray}
where the modifications $P(x), Y_{\mu} (x)$ and $Z_{\mu} (x)$ in
(\ref{modif1}) are assumed to be of order $\theta$ or
higher and such that (\ref{req1}) holds. Furthermore, since
$\theta^{\mu\nu} (x)$ is a local function,
we have to allow for the possibility that it may transform under a
gauge transformation and recognize that, in such a case,
\begin{equation}
\delta \left(A\star B\right) \neq (\delta A)\star B + A\star (\delta
B)\, .
\end{equation}

The analysis is tedious but can be carried out systematically and we find, up
to order $\theta^{2}$, that with
\begin{eqnarray}
D_{\mu}\phi & = & \partial_{\mu}\phi - i A_{\mu}\star \phi +
\frac{1}{2} \partial_{\mu}\theta^{\lambda\rho} \left(A_{\lambda}\star
\partial_{\rho}\phi\right)\nonumber\\
\noalign{\vskip 4pt}%
 &  & - \frac{i}{12}
\partial_{\mu}\left(\theta^{\sigma\tau}\partial_{\sigma}
\theta^{\lambda\rho}\right) \left(\partial_{\tau}A_{\lambda}
\partial_{\rho}\phi -
A_{\lambda}\partial_{\tau}\partial_{\rho}\phi\right) + \frac{1}{12}
\left(2\theta^{\sigma\tau}
\partial_{\mu}\partial_{\sigma}\theta^{\lambda\rho} -
\partial_{\mu}\theta^{\sigma\tau}
\partial_{\sigma}\theta^{\lambda\rho}\right) A_{\lambda}A_{\tau}
\partial_{\rho}\phi\, ,\nonumber\\
\noalign{\vskip 4pt}%
\delta \theta^{\mu\nu} & = & 0\, ,\nonumber\\
\noalign{\vskip 4pt}%
\delta \phi & = & i\epsilon \star \phi\, ,
\nonumber\\
\noalign{\vskip 4pt}%
\delta A_{\mu} & = & \partial_{\mu}\epsilon (x) - i
\left[A_{\mu},\epsilon\right] + \frac{1}{2}
\partial_{\mu}\theta^{\lambda\rho} \left(A_{\lambda}\star
\partial_{\rho}\epsilon\right)\nonumber\\
\noalign{\vskip 4pt}%
 &  & - \frac{i}{12}
\partial_{\mu}\left(\theta^{\sigma\tau}\partial_{\sigma}
\theta^{\lambda\rho}\right) \left(\partial_{\tau}A_{\lambda}
\partial_{\rho}\epsilon -
A_{\lambda}\partial_{\tau}\partial_{\rho}\epsilon\right) + \frac{1}{12}
\left(2\theta^{\sigma\tau}
\partial_{\mu}\partial_{\sigma}\theta^{\lambda\rho} -
\partial_{\mu}\theta^{\sigma\tau}
\partial_{\sigma}\theta^{\lambda\rho}\right) A_{\lambda}A_{\tau}
\partial_{\rho}\epsilon\, ,\nonumber\\
\noalign{\vskip 4pt}%
\delta \left(D_{\mu}\phi\right) & = & i\epsilon\star
\left(D_{\mu}\phi\right)\, ,\label{tfn1}
\end{eqnarray}
the action in (\ref{action1}) is invariant. In this case, the modified
covariant derivative transforms covariantly. There are two main features
worth noting here. First, all the modifications vanish when
$\theta^{\mu\nu}$ is a constant so that (\ref{tfn1}) reduces to
(\ref{tfn0}). The more surprising aspect is that even though
$\theta^{\mu\nu} (x)$ is a local function (and one would naively
expect it to transform in the adjoint representation), it does not
transform under a gauge transformation.

For a real scalar field (matter in the adjoint representation), the
conventional action has the form
\begin{equation}
S_{\rm adj} = \int \mathrm{d} x\, \left(\frac{1}{2} (D_{\mu}\phi)\star
(D^{\mu}\phi) - \frac{m^{2}}{2} \phi\star \phi\right)\,
.\label{action2}
\end{equation}
When $\theta^{\mu\nu}$ is constant, with \cite{hayakawa}
\begin{equation}
D_{\mu}\phi = \partial_{\mu} - i \left[A_{\mu},\phi\right]\, ,
\end{equation}
the gauge transformations
\begin{equation}
\delta \phi = -i\left[\phi,\epsilon\right],\qquad \delta A_{\mu} =
\partial_{\mu}\epsilon (x) - i \left[A_{\mu},\epsilon\right]\, ,
\end{equation}
define an invariance of (\ref{action2}). In generalizing the covariant
derivative and the gauge transformation to the case when
$\theta^{\mu\nu}$ is a local function, we note that (\ref{identity1})
is no longer useful because the field variable is real and, as a
result, even if the covariant derivative transforms covariantly, it
does not help and we have to analyze the invariance of the action as a
whole. 

We can modify the covariant derivative as well as the transformation
laws along the lines of (\ref{modif1}) and write
\begin{equation}
\delta \phi = - i \left[\phi,\epsilon\right] + P (x),\qquad
D_{\mu}\phi = \partial_{\mu}\phi (x) - i \left[A_{\mu},\phi\right] +
Z_{\mu}\, ,\label{modif2}
\end{equation}
with $\delta\theta^{\mu\nu}, \delta A_{\mu}$ already determined in
(\ref{tfn1}). The analysis of the invariance of the action then
determines
\begin{eqnarray}
\delta \phi & = & - i \left[\phi,\epsilon\right]\, ,\nonumber\\
\noalign{\vskip 4pt}%
D_{\mu} \phi & = & \partial_{\mu}\phi - i \left[A_{\mu},\phi\right] +
\partial_{\mu}\theta^{\lambda\rho}
A_{\lambda}\partial_{\rho}\phi\nonumber\\
\noalign{\vskip 4pt}%
 &  & - \frac{1}{2}
\theta^{\sigma\tau}\partial_{\mu}\theta^{\lambda\rho}
\partial_{\sigma}A_{\lambda} A_{\rho} \partial_{\tau}\phi +
\frac{1}{2}
\left(\theta^{\sigma\tau}\partial_{\mu}\partial_{\sigma}\theta^{\lambda\rho} -
\partial_{\mu}\theta^{\sigma\tau}
\partial_{\sigma}\theta^{\lambda\rho}\right) A_{\lambda}
A_{\tau}\partial_{\rho}\phi\, .\label{tfn2}
\end{eqnarray}
It is worth noting here that in determining the invariance of the
action for the scalar field in the adjoint representation, we require
that the  parameter of anti-commutativity be divergenceless, namely,
\begin{equation}
\partial_{\mu}\theta^{\mu\nu} (x) = 0\, ,\label{divergenceless}
\end{equation}
in addition to satisfying the Jacobi identity (\ref{jacobi}). This is a
sufficient condition for the Jacobi identity and is essential in the
discussion of the Seiberg-Witten map in the next section. We point out
here that it is possible, in principle, to
have an invariant action involving modified covariant derivatives and
transformations without using (\ref{divergenceless}), but such
modified quantities become highly non-local as the order of $\theta$
increases and we do not find that very appealing. We also note that
all the modifications in (\ref{tfn2}) vanish when $\theta^{\mu\nu}$ is
constant.

With the construction of the invariant actions for the matter fields,
let us next construct the gauge invariant action for the Maxwell
theory. Conventionally, the invariant action has the form
\begin{equation}
S_{\rm Maxwell} = - \frac{1}{4} \int \mathrm{d} x\, F_{\mu\nu}\star
F^{\mu\nu}\, .\label{action3}
\end{equation}
When $\theta^{\mu\nu}$ is constant, the field strength tensor is given
by
\begin{equation}
F_{\mu\nu} = \partial_{\mu}A_{\nu} - \partial_{\nu}A_{\mu} - i
\left[A_{\mu},A_{\nu}\right] = \left[D_{\mu},D_{\nu}\right]\,
,\label{fieldstrength}
\end{equation}
where $D_{\mu}$ is the covariant derivative (\ref{covariant0}) in the
fundamental representation.  When $\theta^{\mu\nu}$ is local, the
commutator (in the star product sense) of the covariant derivative in
(\ref{tfn1}) does not even give a multiplicative operator so that the
definition of the field strength as well as the analysis of the
invariance of the action (\ref{action3}) need to be done
independently. This is lot more tedious than that for the action for
the matter fields because of the Lorentz structures and we will
present an invariant action up to order $\theta$ although the
procedure can be carried out to any order in $\theta$ in principle.

Here, as in the case of the actions for the matter fields, the idea is
to modify the field strength (\ref{fieldstrength}) as
\begin{equation}
F_{\mu\nu} = \partial_{\mu}A_{\nu} - \partial_{\nu} A_{\mu} - i
\left[A_{\mu},A_{\nu}\right] + X_{\mu\nu} (x)\, ,\label{modif3}
\end{equation}
where $X_{\mu\nu} = - X_{\nu\mu}$ is at least of order $\theta$ and is
determined so that the action
(\ref{action3}) is invariant under the gauge transformations (for the
$\theta^{\mu\nu}, A_{\mu}$) determined in (\ref{tfn1}). To order
$\theta$, this is easily carried out and if we do not assume
(\ref{divergenceless}), the field strength becomes non-local (more so
with increasing order of $\theta$). Therefore, we assume
(\ref{divergenceless}) in which case the modified field strength that
leads to an invariant action has the form
\begin{equation}
F_{\mu\nu} = \partial_{\mu} A_{\nu} - \partial_{\nu} A_{\mu} - i
\left[A_{\mu},A_{\nu}\right] +
\frac{1}{2} \left(\partial_{\mu}\theta^{\lambda\rho}
A_{\lambda}\left(\partial_{\rho} A_{\nu}+F_{\rho\nu}\right) -
(\mu\leftrightarrow \nu)\right)\, .\label{tfn3}
\end{equation}
Once again, we see that the modification is such that it vanishes when
$\theta^{\mu\nu}$ is constant. This demonstrates that there is a
systematic procedure for determining an action (both for matter as well as
gauge fields) which is invariant under $U_{\star} (1)$ transformations
when $\theta^{\mu\nu}$ is a local function.

\section{Seiberg-Witten map}

As in the case when $\theta^{\mu\nu}$ is constant, here we can also
ask if there is a Seiberg-Witten map \cite{witten} that would take the
gauge theory on
the non-commutative manifold to a theory on a commutative space. If we
denote quantities on the non-commutative manifold with ``hats'', then
we wish to determine if there exist functions
\begin{equation}
\hat{\epsilon} = \hat{\epsilon} (\epsilon, A),\qquad \hat{A}_{\mu} =
\hat{A}_{\mu} (A)\, ,\label{map}
\end{equation}
such that
\begin{equation}
\hat{A}_{\mu} (A + \delta_{\epsilon} A) = \hat{A}_{\mu} (A) +
\hat{\delta}_{\hat{\epsilon}}\,\hat{A}_{\mu} (A)\, ,\label{req2}
\end{equation}
where $\delta_{\epsilon} A_{\mu}$ represents the usual $U(1)$ gauge
transformation in a commutative manifold. Namely, we wish to determine
local maps (\ref{map}) in powers of
$\theta$ such that the gauge field in the non-commutative manifold is
(gauge) equivalent to the one on the commutative manifold. We will
determine this to linear order in $\theta$ as is also done in the case
when $\theta^{\mu\nu}$ is constant \cite{witten}, but the procedure
can be carried out to any order in $\theta$.

We recognize that $A_{\mu}$ denotes a $U(1)$ gauge field in a
commutative space so that
\begin{equation}
\delta_{\epsilon} A_{\mu} = \partial_{\mu} \epsilon (x)\,
.\label{tfn4}
\end{equation}
Therefore, using the transformation for the gauge field in
(\ref{tfn1}) and (\ref{tfn4}) in (\ref{req2}) (up to order $\theta$),
we obtain 
\begin{equation}
\hat{A}_{\mu} (A + \partial \epsilon) - \hat{A}_{\mu} (A) -
\partial_{\mu}\hat{\epsilon} -
\theta^{\lambda\rho}\partial_{\lambda}\hat{A}_{\mu} (A)
\partial_{\rho}\hat{\epsilon} - \frac{1}{2}
\partial_{\mu}\theta^{\lambda\rho} \hat{A}_{\lambda}
\partial_{\rho}\hat{\epsilon} = 0\, .\label{req3}
\end{equation}
Let us assume that
\begin{equation}
\hat{A}_{\mu} (A) = A_{\mu} + A'_{\mu} (A),\qquad \hat{\epsilon} =
\epsilon + \epsilon' (\epsilon, A)\, ,
\end{equation}
where the primed quantities are (at least) of order $\theta$. Then,
(\ref{req3}) can be easily solved to determine
\begin{equation}
\hat{A}_{\mu} = A_{\mu} - \frac{1}{2} \theta^{\lambda\rho} A_{\lambda}
(\partial_{\rho} A_{\mu} + F_{\rho\mu}),\qquad \hat{\epsilon} =
\epsilon - \frac{1}{2} \theta^{\lambda\rho} A_{\lambda}
\partial_{\rho}\epsilon\, .\label{map2}
\end{equation}
We recognize this to be exactly the Seiberg-Witten map for the case when
$\theta^{\mu\nu}$ is constant and it continues to hold even in the
case when $\theta^{\mu\nu}$ is a local function. It is interesting
that the extra modifications depending on derivatives of $\theta$ do
identically cancel out so that the usual Seiberg-Witten map holds. In
fact, what is even more interesting is that the field
strength tensor defined in (\ref{tfn3}) goes over under this map to
(up to order $\theta$)
\begin{eqnarray}
\hat{F}_{\mu\nu} & = & \partial_{\mu}\hat{A}_{\nu} -
\partial_{\nu}\hat{A}_{\mu} - i
\left[\hat{A}_{\mu},\hat{A}_{\nu}\right] +
\frac{1}{2}\left(\partial_{\mu}\theta^{\lambda\rho} \hat{A}_{\lambda}
(\partial_{\rho} \hat{A}_{\nu} + \hat{F}_{\rho\nu}) - (\mu\leftrightarrow
\nu)\right)\nonumber\\
\noalign{\vskip 4pt}%
 & = & F_{\mu\nu} - \theta^{\lambda\rho} (F_{\mu\lambda} F_{\rho\nu} +
A_{\lambda} \partial_{\rho} F_{\mu\nu})\, ,\label{map3}
\end{eqnarray}
which represents the same map of the field strength as in the case of constant
$\theta^{\mu\nu}$, even though the field strength in the
non-commutative space in the present case has a much more complicated
structure (\ref{tfn3}). This, therefore, suggests a deeper meaning
underlying the Seiberg-Witten map that deserves further study. We wish
to point out here that all of this works only when
(\ref{divergenceless}) holds.

With the Seiberg-Witten maps determined, we can now easily show to
linear order in $\theta$ that the action for the gauge field
(\ref{action3}) goes over to
\begin{equation}
\hat{S}_{\rm Maxwell} = - \frac{1}{4} \int \mathrm{d}x\,
\hat{F}_{\mu\nu}\star \hat{F}^{\mu\nu} = - \frac{1}{4} \int
\mathrm{d}x\left[\left(1 -
  \frac{1}{2}\theta^{\lambda\rho}F_{\lambda\rho}\right)
  F_{\mu\nu}F^{\mu\nu} + 2 {\rm Tr}\, \theta F^{3} -
  \partial_{\rho}\left(\theta^{\lambda\rho} A_{\lambda}
  F_{\mu\nu}F^{\mu\nu}\right)\right]\, ,\label{action4}
\end{equation}
where we have used an obvious matrix notation in writing the
action. In Eq. (\ref{action4}), we have kept a total divergence which
does not 
contribute to the equations of motion, but is essential for the
definition of the stress tensor of the theory \cite{frenkel,das1}. The
equation of motion
following from the non-commutative theory (to linear order in
$\theta$) has the form
\begin{equation}
\partial_{\mu} \hat{F}^{\mu\nu} + \theta^{\lambda\rho}
\partial_{\lambda}\hat{A}_{\mu} \partial_{\rho} \hat{F}^{\mu\nu} -
\frac{1}{2} \partial_{\mu}\theta^{\nu\rho}
(\partial_{\rho}\hat{A}_{\lambda} +
\hat{F}_{\rho\lambda})\hat{F}^{\mu\lambda} + \partial_{\rho}
\left(\partial_{\mu}\theta^{\lambda\rho} \hat{A}_{\lambda}
\hat{F}^{\mu\nu}\right) - \frac{1}{2}
\partial_{\rho}\left(\partial_{\mu}\theta^{\lambda\nu}
\hat{A}_{\lambda} \hat{F}^{\mu\rho}\right) = 0\, .\label{eqn1}
\end{equation}
On the other hand, the equation of motion following from the mapped
theory in (\ref{action4}) leads to
\begin{equation}
\partial_{\mu}\left[\left(1 - \frac{1}{2} \theta^{\lambda\rho}
  F_{\lambda\rho}\right) F^{\mu\nu} - \left(F^{2}\theta + F\theta F +
  \theta F^{2}\right)^{\mu\nu} - \frac{1}{4} \theta^{\mu\nu}
  F_{\lambda\rho} F^{\lambda\rho}\right] = 0\, .\label{eqn2}
\end{equation}
At first sight, (\ref{eqn1}) does not seem to map into (\ref{eqn2})
under (\ref{map2}) and (\ref{map3}). However, as is the case for the
constant $\theta^{\mu\nu}$ case \cite{frenkel}, the two equations are
identical  under
the map if we use the identity (which holds to order $\theta$)
\begin{equation}
\partial_{\mu} \left[\left(F^{2}\theta\right)^{\mu\nu} + \frac{1}{4}
  \theta^{\mu\nu} F_{\lambda\rho} F^{\lambda\rho}\right] -
  \partial_{\mu}\theta^{\lambda\nu}
  \left(F^{2}\right)^{\mu}_{\;\lambda} = 0\, .
\end{equation}

In a similar manner, we can determine the stress tensor from the
theory in the non-commutative space as well as from the theory
transformed under the Seiberg-Witten map. With the total divergence in
(\ref{action4}), it is straightforward to show that they coincide and
have the form
\begin{eqnarray}
T^{\mu\nu} & = &  \left(1-\frac{1}{2}\theta^{\lambda\rho}
  F_{\lambda\rho}\right) \left((F^{2})^{\mu\nu} +
  \frac{1}{4}\eta^{\mu\nu} F_{\lambda\rho}F^{\lambda\rho}\right) -
  \left((F^{2}\theta F)^{\mu\nu} + (F\theta F^{2})^{\mu\nu} -
  \frac{1}{2} \eta^{\mu\nu} {\rm Tr}\, \theta
  F^{3}\right)\nonumber\\
\noalign{\vskip 4pt}%
 &  & \qquad -
  \partial_{\tau}\left(\theta^{\sigma\tau} A_{\sigma}
  \left((F^{2})^{\mu\nu} + \frac{1}{4}\eta^{\mu\nu} F_{\lambda\rho}
  F^{\lambda\rho}\right)\right)\, .\label{stress}
\end{eqnarray}
The stress tensor is manifestly symmetric and traceless. Using the
equation of motion (\ref{eqn1}) (or (\ref{eqn2})), it can be checked
that this is not, however, conserved as is also the case when
$\theta^{\mu\nu}$ is constant. In that case, we can define a modified
stress tensor that is neither symmetric nor traceless, but
conserved \cite{frenkel}. In contrast, in the present case we find
that even a modified stress tensor such as
\begin{equation}
\overline{T}^{\mu\nu} = T^{\mu\nu} + \partial_{\rho}
\left(\theta^{\lambda\rho} A_{\lambda} T^{(0)\,\mu\nu}\right) -
\left(\theta F T^{(0)}\right)^{\mu\nu}\, ,
\end{equation}
is no longer conserved. Here $T^{(0)}$ denotes the stress tensor
independent of $\theta$. In fact, the divergence of the modified stress
tensor leads to
\begin{equation}
\partial_{\mu} \overline{T}^{\mu\nu} = - \frac{1}{2}
\partial^{\nu}\theta^{\lambda\rho} \left(F
T^{(0)}\right)_{\lambda\rho} = - \partial^{\nu}\theta^{\lambda\rho}
\frac{\delta \hat{S}_{\rm Maxwell}}{\delta \theta^{\lambda\rho}}\,
.\label{conservation}
\end{equation}
We note from (\ref{conservation}) that for a constant
$\theta^{\mu\nu}$, the modified stress tensor would be conserved (even
though it is neither symmetric nor traceless).

The non-conservation of the stress tensor is not hard to
understand. When $\theta^{\mu\nu}$ is a local function, we can think
of the action (\ref{action4}) as representing the interaction (in
addition to the self-interactions) of the
Maxwell field with an external field $\theta^{\mu\nu} (x)$
 and as a result, our system cannot
be thought of as a closed system. It is, of course, not necessary for
a system that is not closed to have conservation of energy. If we have a
complete theory where $\theta^{\mu\nu}$ is a fundamental dynamical
field  (one may
speculate that such a situation may arise in a gravitational theory
with a dynamical $\theta^{\mu\nu}$ as is the case in string theory)
leading to a closed system,
then, the complete energy including that of the dynamical field
$\theta^{\mu\nu}$ has to be conserved. The non-conservation can be
understood yet in a different manner. We recall that conservation of
stress tensor follows from translation invariance of a system. In the
presence of an external field, translation invariance does not hold
(since the external field does not transform). This is manifest in the
last equality in (\ref{conservation}). This brings out a very
interesting feature that contrasts with the case of constant
$\theta^{\mu\nu}$. Namely, it is well known for constant
$\theta^{\mu\nu}$ that translations form a subgroup of the $U_{\star}
(1)$ transformation group \cite{douglas1}. In contrast, when
$\theta^{\mu\nu}$ is
local, we have noted in (\ref{tfn1}) that the parameter of
non-commutativity  does not transform under
a gauge transformation. On the other hand, under translations
$\theta^{\mu\nu}$ has to transform so that we conclude translations do
not form a subgroup of the $U_{\star} (1)$ gauge transformations when
the parameter of non-commutativity is a local function. In retrospect,
in view of the inequality in (\ref{derivative}), it is
clear that when
$\theta^{\mu\nu}$ is a local function, we can no longer represent a
gauge transformation to include translations \cite{douglas1}.

\section{Conclusion}

In this paper, we have analyzed systematically the question of
$U_{\star} (1)$ gauge invariance in a flat non-commutative manifold
where the parameter of non-commutativity is a local function
satisfying the Jacobi identity (and, therefore, leading to an
associative Kontsevich product). We have shown that in this case, the
definitions for  both the covariant derivative as well as the gauge
transformation have to modify in order to have an invariant
action. The modifications can be systematically determined. We have
demonstrated this up to order $\theta^{2}$ in the matter sector (for
both fundamental and adjoint representations) and up to order
$\theta$ in the Maxwell theory. One of the surprising features of this
analysis is that $\theta^{\mu\nu} (x)$ does not transform under a
gauge transformation. The modifications in other variables vanish in the case
when $\theta^{\mu\nu}$ is constant. We have shown that when
$\theta^{\mu\nu}$ is a local function, there exists a Seiberg-Witten
map, which surprisingly coincides with the one for the case when
$\theta^{\mu\nu}$ is constant. We have shown that the equations of
motion as well as the stress tensor in the non-commutative theory go
over under the map to the ones derived from the theory on the
commutative manifold. The stress tensor in the present case is not
conserved and this is traced to non-invariance under translations in
such a theory. We have shown that, unlike the case when
$\theta^{\mu\nu}$ is constant, in the present case translations do not
form a subgroup of the $U_{\star} (1)$ gauge group.

\vskip 1cm

\noindent{\bf Acknowledgment:}

One of us (JF) would like to thank F. T. Brandt for discussions. This work was
supported in part by US DOE Grant number DE-FG 02-91ER40685, by CNPq
and FAPESP, Brazil.

\end{document}